\documentclass[twocolumn]{aastex701}
\usepackage{xcolor} 
\usepackage{amsmath} 
\usepackage{enumitem}

\newcommand{\logg}{\ensuremath{\log\,g}}
\newcommand{\dex}{\ensuremath{\mathrm{dex}}}

\def\spose#1{\hbox to 0pt{#1\hss}}
\def\lta{\mathrel{\spose{\lower 3pt\hbox{$\mathchar"218$}}
     \raise 2.0pt\hbox{$\mathchar"13C$}}}
\def\gta{\mathrel{\spose{\lower 3pt\hbox{$\mathchar"218$}}
    \raise 2.0pt\hbox{$\mathchar"13E$}}}

\received{--}
\revised{--}
\accepted{--}

\shorttitle{A MAGIC Investigation of the Jet Stream}
\shortauthors{Do et al.}

\usepackage[normalem]{ulem}

\begin{document}

\title{The DECam MAGIC Survey: Investigating the Jet Stellar Stream with Photometric Metallicities}

\correspondingauthor{Ha Do}
\email{hado@uchicago.edu}

\author[0009-0009-3497-8369]{H. Q. Do}
\affil{Department of Astronomy \& Astrophysics, University of Chicago, Chicago, IL 60637, USA}
\email{hado@uchicago.edu}

\author[0000-0002-7155-679X]{A. Chiti}
\affil{Kavli Institute for Particle Astrophysics \& Cosmology, Stanford University, Stanford, CA 94305, USA}
\email{achiti@stanford.edu}

\author[0000-0001-6957-1627]{P. S. Ferguson}
\affil{DiRAC Institute, Department of Astronomy, University of Washington, 3910 15th Ave NE, Seattle, WA, 98195, USA.}
\email{pferguso@uw.edu}

\author[0000-0002-4863-8842]{A. P. Ji}
\affil{Department of Astronomy \& Astrophysics, University of Chicago, Chicago, IL 60637, USA}
\affil{Kavli Institute for Cosmological Physics, University of Chicago, Chicago, IL 60637, USA}
\affil{NSF-Simons AI Institute for the Sky (SkAI),172 E. Chestnut St., Chicago, IL 60611, USA}
\email{alexji@uchicago.edu}

\author[0000-0002-9269-8287]{G. Limberg}
\affil{Department of Astronomy \& Astrophysics, University of Chicago, Chicago, IL 60637, USA}
\affil{Kavli Institute for Cosmological Physics, University of Chicago, Chicago, IL 60637, USA}
\email{limberg@uchicago.edu}

\author[0000-0001-9649-8103]{K. R. Atzberger}
\affil{Department of Astronomy, University of University of Virginia, Charlottesville, VA 22903, USA}
\email{katzberger@email.virginia.edu}

\author[0000-0002-3936-9628]{J. L. Carlin}
\affiliation{NSF NOIRLab, Tucson, AZ 85719, USA}
\email{jeffreylcarlin@gmail.com}

\author[0000-0003-1697-7062]{W. Cerny}
\affiliation{Department of Astronomy, Yale University, New Haven, CT 06520, USA}
\email{william.cerny@yale.edu}

\author[0000-0002-9269-8287]{A. Drlica-Wagner}
\affil{Fermi National Accelerator Laboratory, Batavia, IL 60510, USA}
\affil{Department of Astronomy \& Astrophysics, University of Chicago, Chicago, IL 60637, USA}
\affil{Kavli Institute for Cosmological Physics, University of Chicago, Chicago, IL 60637, USA}
\affil{NSF-Simons AI Institute for the Sky (SkAI),172 E. Chestnut St., Chicago, IL 60611, USA}
\email{kadrlica@uchicago.edu}

\author[0000-0003-3081-9319]{G. F. Lewis}
\affiliation{Sydney Institute for Astronomy, School of Physics A28, The University of Sydney, NSW 2006, Australia}
\email{geraint.lewis@sydney.edu.au}

\author[0000-0002-9110-6163]{T. S. Li}
\affiliation{David A. Dunlap Department of Astronomy \& Astrophysics, University of Toronto, 50 St George Street, Toronto ON M5S 3H4, Canada}
\affiliation{Dunlap Institute for Astronomy \& Astrophysics, University of Toronto, 50 St George Street, Toronto, ON M5S 3H4, Canada}
\email{ting.li@astro.utoronto.ca}

\author[0000-0002-3430-4163]{S.~L.~Martell}
\affiliation{School of Physics, University of New South Wales, Sydney, NSW 2052, Australia}
\email{s.martell@unsw.edu.au}

\author[0000-0002-9144-7726]{C.~E.~Mart\'inez-V\'azquez}
\affiliation{NSF NOIRLab, 670 N. A'ohoku Place, Hilo, Hawai'i, 96720, USA}
\email{clara.martinez@noirlab.edu}

\author[0000-0003-0105-9576]{G.~E.~Medina}
\affiliation{David A. Dunlap Department of Astronomy \& Astrophysics, University of Toronto, 50 St George Street, Toronto ON M5S 3H4, Canada}
\affiliation{Dunlap Institute for Astronomy \& Astrophysics, University of Toronto, 50 St George Street, Toronto, ON M5S 3H4, Canada}
\email{gustavo.medina@utoronto.ca}

\author[0000-0002-8282-469X]{N. E. D. Noël}
\affiliation{Department of Physics, University of Surrey, Guildford GU2 7XH, UK}
\email{n.noel@surrey.ac.uk}

\author[0000-0002-6021-8760]{A. B. Pace}
\affil{Department of Astronomy, University of University of Virginia, Charlottesville, VA 22903, USA}
\email{pvpace1@gmail.com}

\author[0000-0003-4479-1265]{V. M. Placco}
\affil{NSF NOIRLab, Tucson, AZ 85719, USA}
\email{vinicius.placco@noirlab.edu}

\author[0000-0001-5805-5766]{A.~H.~Riley}
\affil{Lund Observatory, Division of Astrophysics, Department of Physics, Lund University, SE-221 00 Lund, Sweden}
\email{alexander.riley@fysik.lu.se}

\author[0000-0003-4102-380X]{D.~J.~Sand}
\affil{Steward Observatory, University of Arizona, Tucson, AZ 85721, USA}
\email{dsand@arizona.edu}

\author[0000-0003-1479-3059]{G.~S.~Stringfellow}
\affil{Center for Astrophysics and Space Astronomy, University of Colorado Boulder, Boulder, CO 80309, USA}
\email{Guy.Stringfellow@colorado.edu}

\author[0000-0002-3690-105X]{J.~A.~Carballo-Bello}
\affiliation{Instituto de Alta Investigaci\'on, Universidad de Tarapac\'a, Casilla 7D, Arica, Chile}
\email{jcarballo@academicos.uta.cl}

\author[0000-0001-8536-0547]{L.~R.~Cullinane}
\affiliation{Leibniz-Institut f{\"u}r Astrophysik Potsdam (AIP), An der Sternwarte 16, D-14482 Potsdam, Germany}
\email{lara.cullinane@uq.net.au}

\author[0000-0002-8448-5505]{D.~Erkal}
\affiliation{School of Mathematics and Physics, University of Surrey, Guildford GU2 7XH, UK}
\email{d.erkal@surrey.ac.uk}

\author[0000-0003-2644-135X]{S.~E.~Koposov}
\affiliation{Institute for Astronomy, University of Edinburgh, Royal Observatory, Blackford Hill, Edinburgh EH9 3HJ, UK}
\affil{Institute of Astronomy, University of Cambridge, Madingley Road, Cambridge CB3 0HA, UK}
\email{sergey.koposov@ed.ac.uk}

\author[0000-0001-9649-4815]{B.~Mutlu-Pakdil}
\affiliation{Department of Physics and Astronomy, Dartmouth College, Hanover, NH 03755, USA}
\email{Burcin.Mutlu-Pakdil@dartmouth.edu}

\author[0000-0001-9438-5228]{M. Navabi}
\affiliation{Department of Physics, University of Surrey, Guildford GU2 7XH, UK}
\email{m.navabi@surrey.ac.uk}

\author[0000-0003-2497-091X]{N.~Shipp}
\affil{DiRAC Institute, Department of Astronomy, University of Washington, 3910 15th Ave NE, Seattle, WA, 98195, USA.}
\email{nshipp@uw.edu}

\author[0000-0003-4341-6172]{A.~K.~Vivas}
\affil{Cerro Tololo Inter-American Observatory/NSF NOIRLab, Casilla 603, La Serena, Chile}
\email{kathy.vivas@noirlab.edu}

\author[0000-0001-6455-9135]{A.~Zenteno}
\affil{Cerro Tololo Inter-American Observatory/NSF NOIRLab, Casilla 603, La Serena, Chile}
\email{alfredo.zenteno@noirlab.edu}

\author[0000-0001-6455-9135]{D.~Zucker}
\affil{School of Mathematical and Physical Sciences, Macquarie University, Sydney, NSW 2109, Australia}
\affil{Macquarie University Research Centre for Astronomy, Astrophysics \& Astrophotonics, Sydney, NSW 2109, Australia}
\email{daniel.zucker@mq.edu.au}

\collaboration{30}{(MAGIC \& DELVE Collaborations)}

\begin{abstract}

Stellar streams are dynamically fragile structures formed by the tidal disruption of dwarf galaxies and stellar clusters. These objects are valuable tracers of the gravitational potential and accretion history of the Milky Way, and are key probes for the presence and interactions of starless dark matter subhalos. The Jet stream is a $\sim 30^\circ$-long stellar stream that is situated at 30.4 kpc and originates from a disrupted globular cluster.
It consists of metal-poor stars that follow a retrograde orbit, reducing the impulse imparted from the Milky Way bar and making it especially sensitive to gravitational perturbations from dark matter subhalos. This paper investigates the known extent of the Jet stream by leveraging photometric metallicities derived from a narrowband filter centered on the Ca~{\sc{ii}}~H\&K lines at ($\sim$3950\,{\AA}) on the Dark Energy Camera (DECam), as part of the Mapping the Ancient Galaxy in CaHK (MAGIC) survey. The wide field-of-view of DECam enables the efficient derivation of photometric metallicities for stars across the full extent of the stream, allowing for a metallicity-based selection to identify likely members. We demonstrate the efficacy of photometric metallicities in isolating stream members when used with \textit{Gaia} DR3 proper motions, identifying a sample of 213 candidate Jet stream member stars. This then allows for the study of stream morphology, through which we identify a clear fanning of the stream toward the end farther from the Milky Way bar. We provide a list of candidate members, enabling spectroscopic follow-up of the Jet stream to facilitate further studies of its dynamics.

\end{abstract}

\section{Introduction} 
\label{sec:intro}

The dark energy plus cold dark matter ($\Lambda$CDM) cosmological paradigm models the matter content of the universe as being dominated by dark matter, with galaxies forming through the hierarchical merging of smaller structures. This paradigm arose from observations of the large-scale distribution of galaxies (\citealt{ps+74}; \citealt{wr+78}) and of residual signatures from the aforementioned accretion events \citep[e.g.,][]{sz+78}. A local manifestation of this hierarchical assembly process is stellar streams, elongated associations of stars that form when dwarf galaxies or globular clusters are tidally disrupted during infall onto a larger host (e.g., \citealt{lb+95}; \citealt{nc+16}, \citealt{bp+25}). Their physical scale as well as their dynamically fragile structure render them especially sensitive to gravitational perturbations, both from baryonic matter and dark matter (\citealt{jsb+01}; \citealt{c+13}; \citealt{esb+16}; \citealt{bhp+19}; \citealt{bbb+21}). As such, stellar streams are valuable testing grounds for the $\Lambda$CDM model, preserving a fossil record of stellar and structural formation while simultaneously tracing the gravitational field in the dark-matter-dominated outskirts of galactic halos.

The $\Lambda$CDM model also predicts the existence of substantial, small-scale dark matter substructure within larger structures such as galaxy clusters and galactic halos (\citealt{kkv+199}, \citealt{mgg+99}, \citealt{ghs+04}; \citealt{dms+05}; \citealt{wbf+20}). In the case of our Milky Way, thousands of dark matter subhalos with mass greater than $10^6$ solar masses are expected to orbit within the halo \citep{swv+08}. Cold stellar streams, which are particularly sensitive to dynamical perturbations (e.g., fly-by interactions), are promising probes to look for gravitational signatures from these yet-to-be-detected halos. These interactions may alter the density distribution of stellar streams, resulting in features such as gaps or misalignments \citep{jsh+02, ili+02}. In this framework, perturbations in stellar streams at large galactocentric radii, where there are fewer baryonic perturbations from structures such as giant molecular clouds and the Milky Way bar (e.g., \citealt{lke+21}), would be key signatures of the presence of dark matter subhalos in the local galaxy. Studying any detected aberrational features could place limits on the population of dark matter subhalos in the Milky Way, constrain the nature of dark matter, and evaluate the $\Lambda$CDM model \citep{pwb+18}.

Interactions between stellar streams and dark matter subhalos can manifest as small-scale density and velocity perturbations. However, detecting and characterizing these features from broadband photometry alone is exacting due to the large physical extents (e.g.,~$>10^\circ$; \citealt{taa+22}) and low surface brightness of stellar streams ($\geq 28.5$\,mag\,arcsec$^{-2}$; e.g., \citealt{sdb+18}) requiring both deep and wide photometric coverage. Further, contamination from foreground stars and background galaxies pose a significant challenge in discriminating between stream members and contamination. The advent of the \textit{Gaia} astrometric mission \citep{gaia+16, gaia+23} has increased the number of known and characterized stellar streams by an order of magnitude to more than a hundred, as it allows improved member selection using proper motion/parallax information. This availability of proper motion data for brighter stars ($G < 21$) has shown that density variations and velocity structure are common in these streams (\citealt{mi+18}, \citealt{imm+21}, \citealt{m+23}, \citealt{bp+25}). Spectroscopic velocities and metallicities for individual stars, on top of 5D data ($\alpha, \delta, \mu_\alpha, \mu_\delta, d$) would provide the cleanest selection of member stars suitable for the analysis of a stream's morphology, allowing rigorous interpretation of the nature and origin of potential perturbations from dark matter subhalos. 
However, spectroscopic observations are expensive to make, constrained by the number and arrangement of slits/fibers, and limited to brighter stars relative to photometry \citep{taa+22}. 
The costliness of taking spectra for every possible member star motivates the exploration of other methods to obtain a pure sample of member stars across the full extent of a stellar stream.

The Jet stream is a stellar stream that is a particularly promising environment to look for signatures of dark subhalo interactions. The Jet stream was discovered by \citet{jtn+18} as part of the Search for the Leading Arm of Magellanic Satellites (SLAMS) survey and placed at a galactocentric distance of 30.4 kpc by \citet{fsd+22}, based on blue horizontal branch (BHB) stars. The 12.7 kpc pericenter of its orbit \citep{fsd+22} is substantially far from the center of our Galaxy to reduce the likelihood of baryonic perturbations. The progenitor of the stream is posited to be a globular cluster due to its narrow width, though no identifiable progenitor has been found (\citealt{jtn+18}). This suggested globular origin means that the Jet stream ought to be dynamically cold, thin, and especially sensitive to gravitational disruptions. For a stream originating from a globular cluster, the Jet stream consists of a population of particularly metal-poor stars ([Fe/H] = $-2.38\pm0.03$; \citealt{taa+22}). 
It is also one of the few known stellar streams on a retrograde orbit, making it less likely than a prograde stream to have experienced a significant impulse from the Milky Way bar, and dynamical modeling suggests the Large Magellanic Cloud has had little impact on the stream (\citealt{fsd+22}, \citealt{taa+22}). This positions the Jet stream as an excellent candidate for probing interactions with dark subhalos.

This project builds on previous work demonstrating that photometry with filters covering the prominent Ca~{\sc{ii}}~H\&K lines (hereafter CaHK) can be used to derive stellar metallicities (i.e., \citealt{abl+91}, \citealt{ksb+07}, \citealt{smy+17}, \citealt{cfj+20}, \citealt{wpb+21}, \citealt{hyl+22}, \citealt{fws+23}). The Mapping the Ancient Galaxy in CaHK (MAGIC; Chiti et al., in prep) survey, conducted using DECam \citep{fdh+15} with its large 2.2{\degr} field-of-view at the Víctor M. Blanco 4\,m telescope located at the NSF Cerro Tololo Inter-American Observatory (CTIO), offers an efficient method of deriving stellar metallicities across large regions of the sky using a narrowband filter that is sensitive to the CaHK absorption features at 3968.5\,{\AA } and 3933.7\,{\AA}. Especially in the case of stellar streams, as they host stars that are on average more metal-poor than those in the Milky Way, the photometric stellar metallicities derived from this method can be used to distinguish stream members from the foreground population as a first cut \citep[e.g.,][]{apk+26}. 

In this paper, we use data from the DECam MAGIC survey combined with targeted observations of the Jet stream to image the entirety of its known extent with narrow-band CaHK photometry. Specifically, we use photometric metallicities obtained through the CaHK filter in  tandem with broadband $g, i$ photometry from the DECam Local Volume Exploration (DELVE) Survey DR2 \citep{dfa+22} to perform a color-magnitude and metallicity cut to isolate candidate stream members. We also utilize proper motions from \textit{Gaia} DR3 \citep{gaia+21} to identify stream members; in combination, these selections allow us to identify a relatively pure population of stream member stars. Our goal is to test density and width variations, as well as to verify features reported in previous broadband photometric studies of the Jet stream, while also searching for new features. We also aim to gauge the efficacy and reliability of photometric metallicities from DECam as a method of confirming stream membership by comparison with a subset of spectroscopically confirmed stream members from the Southern Sky Stellar Stream Spectroscopic Survey ($S^5$; \citealt{lkz+19}).

This paper is organized as follows. In Section~\ref{sec:obs}, we describe the data catalog and cover the use of photometry in deriving metallicities.
In Section~\ref{sec:analysis}, we describe our analysis of the Jet stream, selecting for stars using distance, proper motion, and metallicity. In Section~\ref{sec:discussion}, we discuss the results of our analysis, including the completeness and purity of the resulting selection of candidate stars, as well as assess stream morphology and implications for further studies.
In Section~\ref{sec:conclusion}, we conclude and summarize our results. 

\section{Observations, Catalog Generation, and Photometric Metallicities}
\label{sec:obs}

\subsection{Observations}
\label{subsec:obs}
\begin{figure*}[!htbp]
\centering
\includegraphics[width =\textwidth]{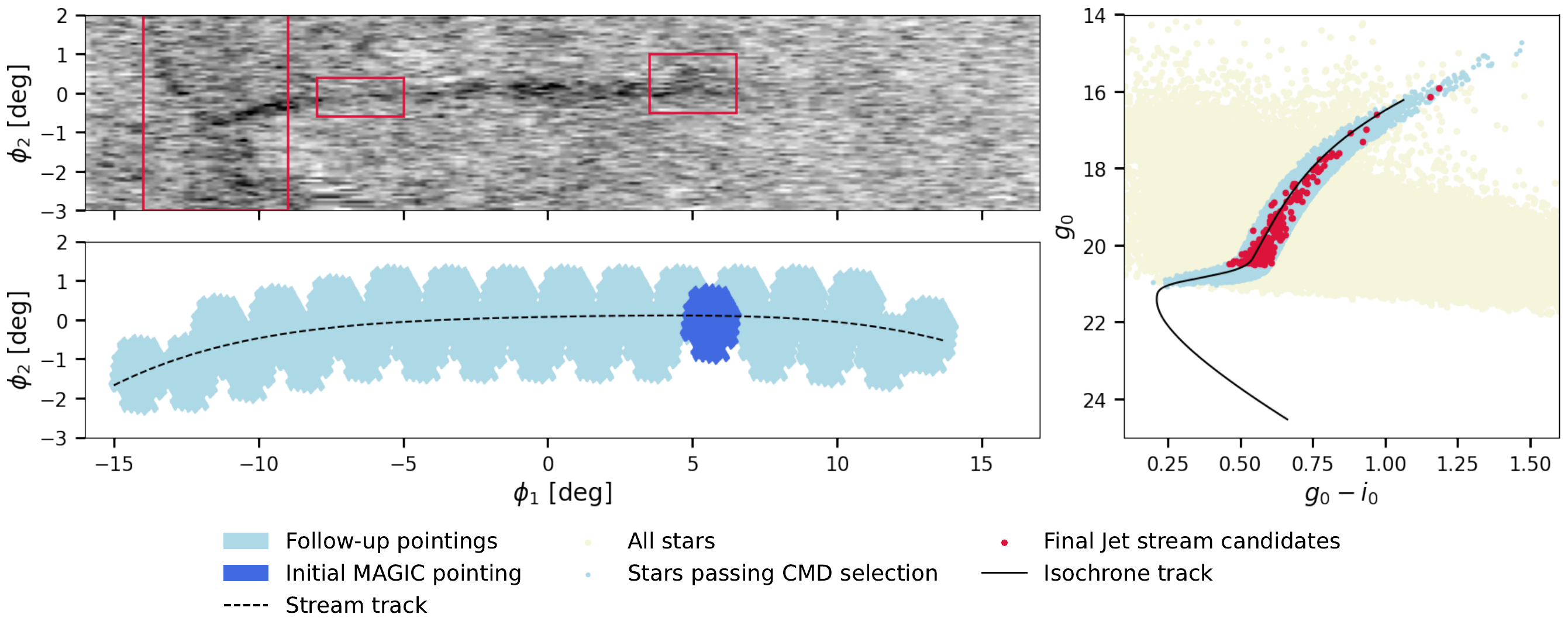}
\caption{Top left: Density map of the Jet stream from \citet{fsd+22}. Tentative perturbation features proposed by \citet{fsd+22} are highlighted in red boxes, including a $\sim 4^\circ$ gap at $\phi_1$ = $-6^\circ$ and a feature at $\phi_1$ = $-12.5^\circ$ (see Section~\ref{sec:analysis} for the definition of stream coordinates) cutting across the stream track. Bottom left: The stream track overlaid on the coverage map of the Jet stream; the initial MAGIC pointing is highlighted in dark blue while all other follow-up 2025A exposures are in light blue. Right: Empirical isochrone ([Fe/H] =$ -2.3$, Age = 12.7; see Section~\ref{subsec:cmd}) used for selecting Jet stream members, with stars passing the color-magnitude diagram (CMD) cut shown in blue. The final sample of member stars, having undergone CMD, proper motion, and metallicity cuts (Section~\ref{sec:analysis}), are highlighted in red.}
\label{fig:footprint}
\end{figure*}

The Jet stream was covered using a combination of one MAGIC field from pilot survey data (NOIRLab Prop. ID 2023A-933926; P.I. Anirudh Chiti) and 26 follow-up fields outside of the MAGIC footprint obtained in March 2025 (NOIRLab Prop. ID 2025A-568024; P.I. Anirudh Chiti \& Ha Do). The data from this aggregate coverage are made available online \citep{zenodo}\footnote{\url{https://doi.org/10.5281/zenodo.19475550}}. In total, these fields cover $\sim30$\,{\degr} lengthwise with a width of $\sim 3.3$\,{\degr}, spanning the complete known length of the Jet stream (\citealt{fsd+22}; Figure~\ref{fig:footprint}).
The data were obtained using DECam and a narrow-band filter designated N395. This filter has a central wavelength of 3951.9\,{\AA} and a bandpass of 100.1\,{\AA}, with an approximate “top hat” transmission profile; designed to capture the CaHK absorption features at 3968.5\,{\AA } and 3933.7\,{\AA} respectively. The original MAGIC pointing had a dithered dark-time exposure of 3 hr (9$\times$20 min), while all extended fields have an exposure time of at least 36 min, obtained through three dithered exposures (3$\times$12\,min) to cover chip gaps. During observations, a number of fields were revisited based on seeing conditions, such that all fields had at least one exposure with a seeing value $\leq$1.2 arcsec. The overall average seeing was $\sim1.0$ arcsec. The CaHK photometry has a minimum signal-to-noise (S/N) value of $\sim$19 at a depth of $g \sim 21.5$ across all fields, and a median S/N $\sim$ 46 at the base of the red giant branch (RGB) ($g = 20.5$ at the distance modulus of the Jet stream $m - M = 17.55$). 
After deriving photometric metallicities (as described in Section~\ref{subsec:FeH}), we find that this photometric precision achieves a median metallicity error of $\sim 0.23$ dex at [Fe/H]$_{\text{CaHK}} = -2.0$, with a maximum metallicity uncertainty of 0.29 dex for sources that pass accompanying color-magnitude and proper motion selections for Jet stream membership down to $g\sim20.5$ (see Section~\ref{sec:analysis}).

\subsection{Photometry \& Catalog Generation}

We retrieved reduced N395 images after they were processed by the DECam Community Pipeline \citep{vgd+14}, and generated source catalogs from individual exposures using \texttt{Source Extractor} and \texttt{PSFEx} \citep{ba+96, b+11, b+13}.
Photometric zeropoint calibration was performed following the prescription used in the MAGIC survey (Chiti et al., in prep.), by cross-matching the N395 instrumental magnitudes to the synthetic narrow-band CaHK magnitudes derived from \textit{Gaia} XP spectra \citep{gaia+16, gaia+21, xp_1, xp_2, xp_3}.
As a brief summary, \textit{Gaia} DR3 included low-resolution ($R \sim$ 25--100) flux-calibrated spectra from 3300\,{\AA} to 10500\,{\AA} for 220 million stars, known as XP spectra, along with a toolkit (\texttt{GaiaXPy}\footnote{https://gaia-dpci.github.io/GaiaXPy-website/}) to derive synthetic photometry from this data in a variety of bandpasses \citep{gaia+21}.
We calculate the flux through the default CaHK bandpass in \texttt{GaiaXPy}, which mimics the filter used in the Pristine survey \citep{smy+17}, on which the DECam N395 filter is based.
We cross-match the resulting XP synthetic CaHK photometry with our N395 instrumental magnitudes to derive a photometric zeropoint for each DECam pointing. 
We note that the DECam N395 filter has a bandpass that is not a pure top-hat due to the throughput of the DECam CCDs, but we approximate the zeropoints as such since the synthetic photometry for photometric metallicities is generated under the assumption of a top-hat filter. 
After zero-point correction, source catalogs from individual pointings are then combined by taking a weighted average of the N395 photometry. 
This combined N395 catalog is then cross-matched to data from DELVE DR2 for broadband DECam $g, r, i$ photometry, and \textit{Gaia} DR3 \citep{gaia+21} for proper motion data to generate a final catalog that includes all of the observables for our analysis. 

\subsection{CaHK Photometric Metallicities}

The primary methodological goal of this study is to use metallicities derived from the DECam CaHK photometry to isolate a clean sample of Jet stream members, given its low metallicity ([Fe/H] = $-2.38$; \citealt{taa+22}) relative to the Milky Way halo population (e.g., mean [Fe/H] $\approx -1.2$ to [Fe/H] $\approx-1.6$; \citealt{cnz+19,ysm+20}). 
In this section, we describe our methodology in deriving photometric metallicities using the narrowband CaHK filter, in tandem with DELVE DR2 broadband $g, i$ photometry. 
The utility of using narrow-band CaHK photometry to derive metallicities or identify low metallicity stars has been demonstrated by a number of studies over the years (e.g., \citealt{ksb+07}, \citealt{smy+17}, \citealt{cfj+20},  \citealt{hyl+22}, \citealt{ash+22}, \citealt{paa+22}), with \citet{bcl+25}, \citet{placco+25}, \citet{pace+25}, \citet{Chiti+25}, and \citet{apk+26} in particular demonstrating the capabilities of the DECam CaHK filter as early results from the MAGIC survey. 
This work makes use of the same approach, which is adapted from the forward-modeling procedure in \citet{cfj+20, cfm+21}, in which synthetic photometry in the CaHK, $g$, and $i$ bandpasses are generated using synthetic spectra spanning a wide range of stellar parameters, and compared to the observed photometry to derive metallicities.

\begin{figure}[!htbp]
\includegraphics[width = \columnwidth]{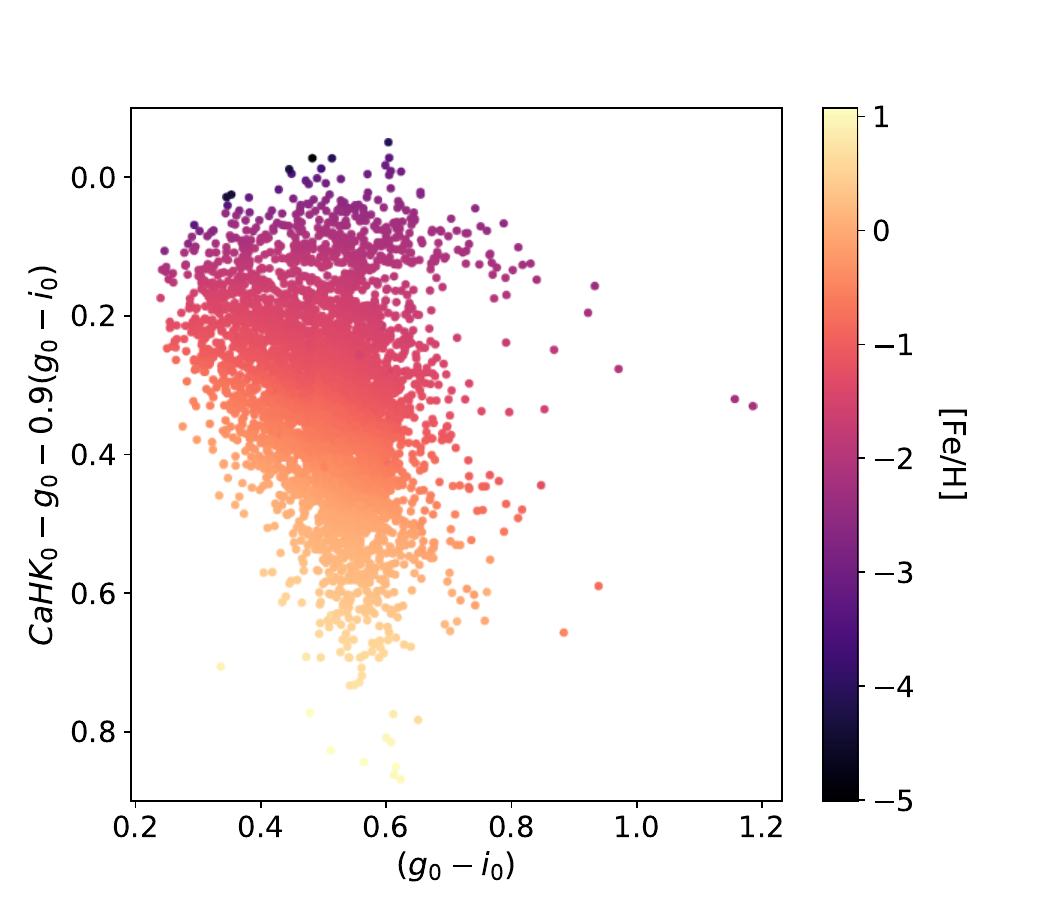}
\caption{Color–color diagram of stars observed in this aggregate Jet stream survey that have undergone CMD and proper motion selection, where each star is colored by the inferred photometric metallicity from the CaHK photometry. 
Stars of different metallicities form distinct bands in this space, enabling photometric separation.}
\label{fig:filter_color}
\end{figure}

Specifically, the grid of synthetic spectra is generated using the Turbospectrum radiative transfer code \citep{ap+98, p+12}, with MARCS model atmospheres \citep{gee+08} and line data from the VALD database \citep{pkr+95, rpk+15} as inputs. 
We generated this grid to span log\,$g$ = $-$0.5\,dex to 5.5\,dex in steps of 0.5\,dex, and metallicities from [Fe/H] = $-$5\,dex to 1\,dex with discrete points at [Fe/H]~=~\{$-5.0, -4.0, -3.0, -2.5, -2.0$, $-1.5, -1.0$, $-0.7$, $-0.5, -0.2, 0, +0.2, +0.5, +0.7, +1.0$\}.
Synthetic photometry in the CaHK and DECam $g,i$ filters were then derived from this grid of flux-calibrated synthetic spectra following Section~3.2 in \citet{cfj+20}, which closely follows the formalism in \citet{cv+14}.
This synthetic $\text{CaHK}_0$, $g$, and $i$ photometry was used to populate 
the metallicity-sensitive color-color space $\text{CaHK}_0 - g_0 + 0.9\times(g_0 - i_0)$ vs. $g_0 - i_0$ (as shown in Figure~\ref{fig:filter_color}).
This space was interpolated with a 2d cubic spline function (\texttt{scipy.interpolate.griddata}; \citealt{vgo+20}) to generate a continuous mapping between each point in this space and a metallicity (see Figure~\ref{fig:filter_color}).


The observed magnitudes were corrected for dust extinction using the \citet{sfd+98} map, and the CaHK extinction coefficient from \citet{smy+17} of $A_{\text{CaHK}}/E(B-V) = 3.924$.
A metallicity was then derived for each star from its location on the $\text{CaHK}_0 - g_0 + 0.9\times(g_0 - i_0)$ vs. $g_0 - i_0$ plot, based on the interpolation function described in the last paragraph.
We note that the values of the interpolation function shift slightly based on the assumed $\logg$.
To account for this, we adopt a two-step iteration approach. 
First, $\logg$ values for each star are adopted by matching its $g_0-i_0$ color to the RGB of 10\,Gyr, Dartmouth isochrones \citep{dcj+08} at metallicities of [Fe/H] = $-1.0, -1.5, -2.0$, and $-2.5$. 
The synthetic photometry contours produced by the nearest $\logg$ value in the synthetic grid are then used to populate the $\text{CaHK}_0 - g_0 + 0.9\times(g_0 - i_0)$ vs. $g_0 - i_0$ space to derive metallicities for the $\logg$ from each isochrone.
These metallicities are then averaged, and the $\logg$ from the isochrone at the closest metallicity to the average is taken as the $\logg$ of the star.
The metallicity derived using that $\logg$ is taken to be the metallicity of the star.
The uncertainty on each metallicity is derived by shifting the photometric values by the error in the CaHK, $g$, and $i$ bands and adding the shifts in the resulting metallicity in quadrature. 
A minimum uncertainty floor of 0.16\,dex is then added in quadrature to these uncertainties, following \citet{cfj+20} and \citet{bcl+25}, who respectively showed that this floor led to total uncertainties that were consistent with metallicities in globular clusters and spectroscopic studies of the Sculptor dwarf galaxy.

\section{Methods and Analysis}
\label{sec:analysis}

In this section, we outline the procedure for the selection of member stars of the Jet stream. We begin by applying preliminary quality cuts to the full survey sample to ensure reliable astrometry, photometry, and derived stellar parameters.

From \textit{Gaia} DR3 astrometry, we require a parallax signal-to-noise ratio $\varpi/\sigma_\varpi < 3$ (\texttt{parallax/parallax\_error} $< 3$), as stars with resolved parallaxes would be too close to be a member of the Jet stream. Additionally, we retain only sources that have broadband colors ($g_0-r_0$ and $r_0-i_0$) consistent with the stellar locus (\texttt{broadband\_valid} $=$ True) and exclude sources that are flagged as variable in the \textit{Gaia} DR3 variable star catalog (\texttt{gaia\_var\_flag} $=$ False; \citealt{gaiavar1}). Finally, we exclude stars with colors outside the bounds of the grid of synthetic \textit{Gaia} XP photometry used to compute photometric metallicities in the \textsc{MAGIC} pipeline \texttt{(feh\_extrapolation\_flag} $= 0$).

We then utilize three main selection criteria in order to select for Jet stream members: a CMD selection, proper motion selection, and photometric metallicity selection. Following \citet{fsd+22}, we first transform the Jet stream coordinates to a stream-centered coordinate frame, defined by the pole \((\alpha_{\text{pole}}, \delta_{\text{pole}}) = (64.983^\circ, 34.747^\circ)\), and a \(\phi_1\) center of \(\phi_1 = 63^\circ\) 
(\(\phi_1, \phi_2 = 0^\circ, 0^\circ\) at \((\alpha, \delta) = (138.62^\circ, 22.10^\circ)\)).
The rotation matrix used to convert \(\alpha, \delta\) to \(\phi_1, \phi_2\) as well as \(\mu_\alpha, \mu_\delta\) to $\mu_{\phi_1}$, $\mu_{\phi_2}$ is the following:
\[
R = 
\begin{bmatrix}
  -0.69798645 & 0.61127501 & -0.37303856 \\
  -0.62615889 & -0.26819784 & 0.73211677 \\
  0.34747655 & 0.74458900 & 0.56995374
\end{bmatrix}
\]

This conversion to $\phi_1$ and $\phi_2$  allows us to use the proper motion and distance relations provided by \citet{fsd+22}. After converting $\alpha$, $\delta$ in our catalog to stream coordinates $\phi_1$, $\phi_2$, we perform a color–magnitude selection using an empirically constructed isochrone (detailed in Section~\ref{subsec:cmd}). We then apply proper motion cuts derived from Monte Carlo propagation of measurement uncertainties from \textit{Gaia} DR3 into the stream frame, ensuring a statistically consistent kinematic selection across the length of the stream. Finally, we incorporate a photometric metallicity selection centered on [Fe/H]$_{\text{CaHK}} = -2.18$, higher than prior spectroscopic determinations ([Fe/H] $= -2.38$ as reported by \citealt{taa+22}; see subsection~\ref{subsec:FeH}). A summary of the parameter values used for this selection process is provided in Table~\ref{tab:params}. These cuts yield a relatively pure (purity discussed in Section~\ref{subsec:purity}) sample of Jet stream candidates for further analysis of its morphology in Section~\ref{sec:discussion}.

\subsection{Color-Magnitude Selection}
\label{subsec:cmd}
All Jet stream stars in our selection lie above the main-sequence turn-off (g $\approx$ 21) as we use \textit{Gaia} DR3 proper motions, which we limit to g $\sim$ 20.5 for relatively low errors; main-sequence stars below the turnoff would be fainter than this magnitude cut. However, we observed a deviation between known member stars in the first data release of the $S^5$ survey and theoretical Dartmouth isochrones \citep{dcj+08} in the RGB. In order to resolve this difference, we instead opt to construct an empirical isochrone based on the stellar population of NGC 5053, a low-mass globular cluster with [Fe/H]~=~$–2.27$ \citep{h+96} and an age of 12.7 Gyr \citep{k+18}, broadly consistent with the properties of the Jet stream. 

\begin{figure*}[!htbp]
\centering
\includegraphics[width =\textwidth]{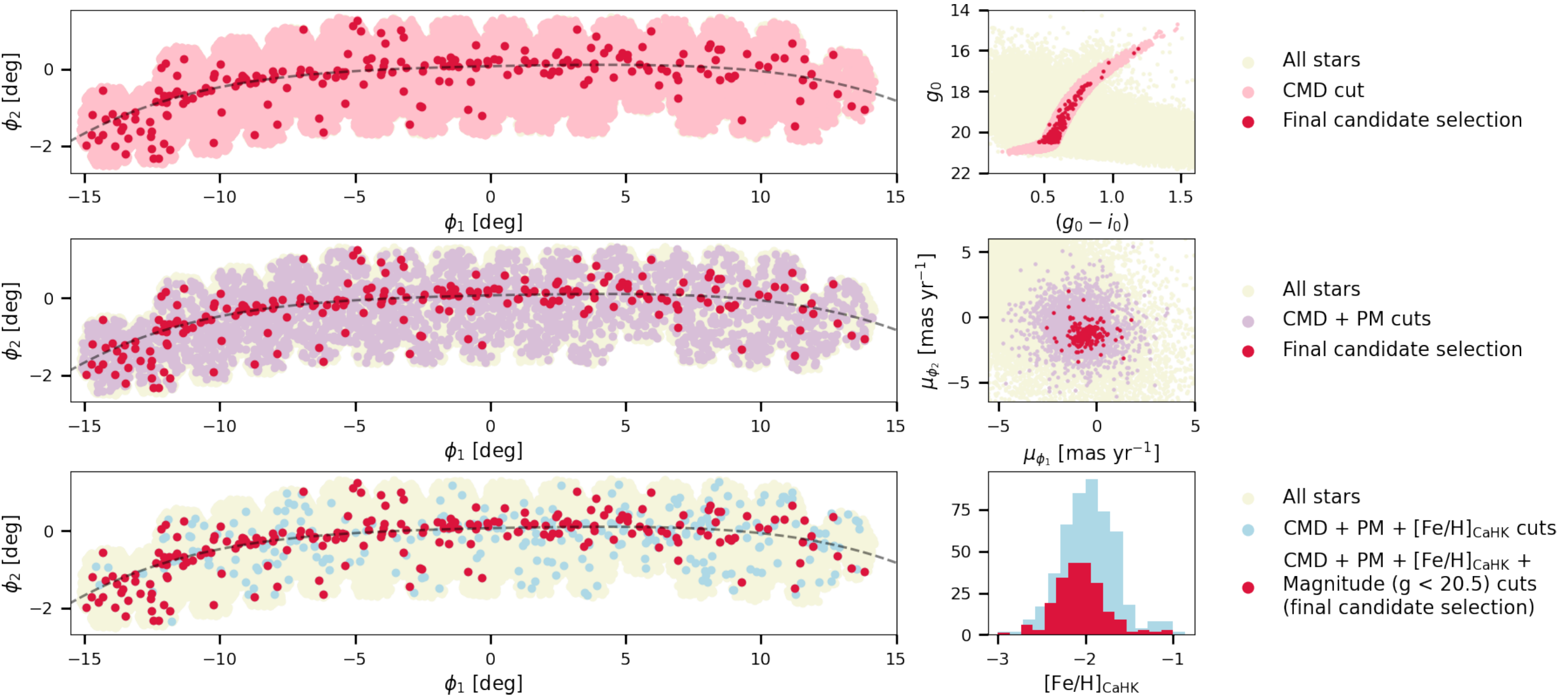}
\caption{Summary of the sequential selection of Jet stream candidate stars based on color–magnitude, proper motion, and metallicity criteria. Beige points in all panels are all observed stars while red points in all panels are the final selection of stars, having passed all three of the aforementioned cuts and with magnitudes $g_0$ brighter than 20.5. Top Left: Spatial distribution in stream coordinates after applying a color–magnitude selection. Pink points denote stars passing the CMD selection. Top Right: CMD showing all stars in the survey region (beige), CMD-selected stars (pink). Middle Left: Same spatial plane, now highlighting stars that pass both the CMD and proper motion selections (purple). Middle Right: Proper motion distribution of stars passing the CMD and proper motion cut (purple), centered around the Jet stream’s mean motion. Bottom Left: Blue data points are stars with photometric metallicities consistent within 2\,$\sigma$ of [Fe/H]$_{CaHK}$ $= -2.18$ and with [Fe/H]$_{CaHK}$ uncertainties $< 0.7$ dex. Red points are stars passing an additional magnitude cut for stars brighter than $g_0 = 20.5$, comprising our final catalog of Jet stream candidates. Bottom Right: Histogram of photometric metallicities for stars having undergone a CMD, proper motion and [Fe/H]$_{CaHK}$ selection (blue) and final candidate stars (red), demonstrating the effectiveness of the metallicity cut in isolating a metal-poor population consistent with the Jet stream.}
\label{fig:cuts}
\end{figure*}

We generate the isochrone by fitting a cubic 1-D spline with internal knots using \texttt{scipy.interpolate.LSQUnivariateSpline} to the stellar population of NGC 5053 in the $g-$band magnitude versus $(g_0 – i_0)$ color using values from DELVE DR2 \citep{dfa+22} (see right panel of Figure~\ref{fig:footprint}). 
The resulting empirical isochrone was used to define a polygon for a  color–magnitude selection following \citet{fsd+22}, allowing a $\pm$0.05\,mag tolerance in $g_0 – i_0$ color around the isochrone, which is then adjusted along the stream track to account for the linear distance gradient in the literature, $\frac{\mathrm{d}\,d_\text{Jet}}{d\,\phi_1} = -0.2$\,kpc\,deg$^{-1}$, as established in \citet{fsd+22}.
We initially applied the selection across a range of distance moduli from 17.3 to 17.9 to the MAGIC field, identifying an optimal match at $\mu$ = 17.55 for the segment of the Jet stream at $\phi_1 = [5^\circ, 7^\circ]$, $\phi_2 = [-1^\circ, 1^\circ]$. Adopting this reference point, we extend the selection across the full length of the Jet stream to account for the Jet stream's distance variation, ensuring consistent selection of candidate members across its full extent using the following function:
\begin{equation}
(m - M) = -(\phi_1 - 5.6) \times 0.014 + 17.55
\end{equation}

\noindent where $\phi_1$ is in degrees.
Accordingly, we derived the expected distance modulus for each star in our catalog based on $\phi_1$, if it were a member of the Jet stream. The empirical isochrone described above was then overlaid at this distance modulus, and stars passed this selection if their location on the CMD was consistent within $\pm$0.05\,mag of this isochrone. The result of this selection is shown in the top panel of Figure~\ref{fig:cuts}, retaining 54,054 of the initial 611,961 observed stars passing initial quality cuts.

\subsection{Proper Motion selection}
Following \citet{fsd+22}, we convert proper motions from \textit{Gaia} DR3 \citep{gaia+21} into the stream-aligned coordinate system $(\mu_{\phi_1}, \mu_{\phi_2})$, where proper motions are in $\mathrm{mas\,yr^{-1}}$ and $\phi_1$ is in degrees. We define functions of $\mu_{\phi_1}$ and $\mu_{\phi_2}$ along $\phi_1$ in order to account for variation in the stream’s kinematics at different locations along the length of the stream. The functions for $\mu_{\phi_1}$ and $\mu_{\phi_2}$ are as follows, respectively:
\begin{equation}
\mu_{\phi_1}(\phi_1) = -0.933- 0.0109\,\phi_1
\end{equation}

\begin{equation}
\mu_{\phi_2}(\phi_1) = -0.023 + 0.0120\,\phi_1
\end{equation}

\noindent where again $\phi_1$ is in units of degrees, and $\mu_{\phi_1}, \mu_{\phi_2}$ are in $\mathrm{mas\,yr^{-1}}$. The uncertainties in this stream proper motion as a function of $\phi_1$ were adopted from \citet{fsd+22}:

\begin{equation}
\sigma_{\mu_{\phi_1}}(\phi_1) = \sqrt{(0.018)^2 + \left(0.0030\,\phi_1 \right)^2}
\end{equation}

\begin{equation}
\sigma_{\mu_{\phi_2}}(\phi_1) = \sqrt{(0.016)^2 + \left( 0.0021\,\phi_1 \right)^2}
\end{equation}

\noindent where $\sigma_{\mu_{\phi_1}}$ and $\sigma_{\mu_{\phi_2}}$ are in units of $\mathrm{mas\,yr^{-1}}$, and $\phi_1$ is in degrees. We then perform a Monte Carlo propagation of the kinematic errors for each star passing the initial CMD selection into stream coordinates, as our matrix formalism for the transformation from equatorial coordinates to stream coordinates does not generate errors. For each object, we construct a covariance matrix in equatorial proper-motion space using \textit{Gaia} DR3 \texttt{pmra\_pmdec\_corr} ($\rho_{\alpha\delta}$):
\begin{equation}
\mathbf{C}_{\alpha\delta} =
\begin{bmatrix}
\sigma_{\mu_{\alpha}}^2 &
\rho_{\alpha\delta}\,\sigma_{\mu_{\alpha}}\,\sigma_{\mu_\delta} \\
\rho_{\alpha\delta}\,\sigma_{\mu_{\alpha}}\,\sigma_{\mu_\delta} &
\sigma_{\mu_\delta}^2
\end{bmatrix},
\end{equation}
and draw 1,000 realizations from the corresponding bivariate Gaussian distribution. Each realization was rotated into ($\mu_{\phi_1},\mu_{\phi_2}$) space using the aforementioned rotation matrix, resulting in the mean proper motion and the associated covariance matrix in stream coordinates. The diagonal elements of the covariance matrix provide the propagated uncertainties in ($\mu_{\phi_1},\mu_{\phi_2}$), while the off-diagonal term captures the correlation in the rotated frame.

To assess the kinematic consistency of each star with the stream model, we constructed the total covariance matrix
\begin{equation}
\mathbf{C} = \mathbf{C}_{obs} + \mathbf{C}_{model} + \mathbf{C}_{intrinsic},
\end{equation}
where $\mathbf{C}_{obs}$ is the Monte Carlo covariance in stream coordinates, $\mathbf{C}_{model}$ is the diagonal matrix formed from the model dispersions at the star’s $\phi_1$, and $\mathbf{C}_{intrinsic}$ is a diagonal matrix representing an adopted intrinsic dispersion value of $\sigma_{\mu, intrinsic}\,\sim\,0.04$ mas yr$^{-1}$ following \citet{fsd+22}.
We then computed the Mahalanobis distance D, using \texttt{scipy.spatial.distance.mahalanobis}. The squared Mahalanobis distance follows a chi-square distribution with two degrees of freedom; as such, in order to select stars within a $\sim\,99\%$ confidence interval, we select for stars with a corresponding Mahalanobis distance D $<2.5$. This cut retains 3,119 stars from the previous selection.

\begin{figure*}[!htbp]
\centering
\includegraphics[width =\textwidth]{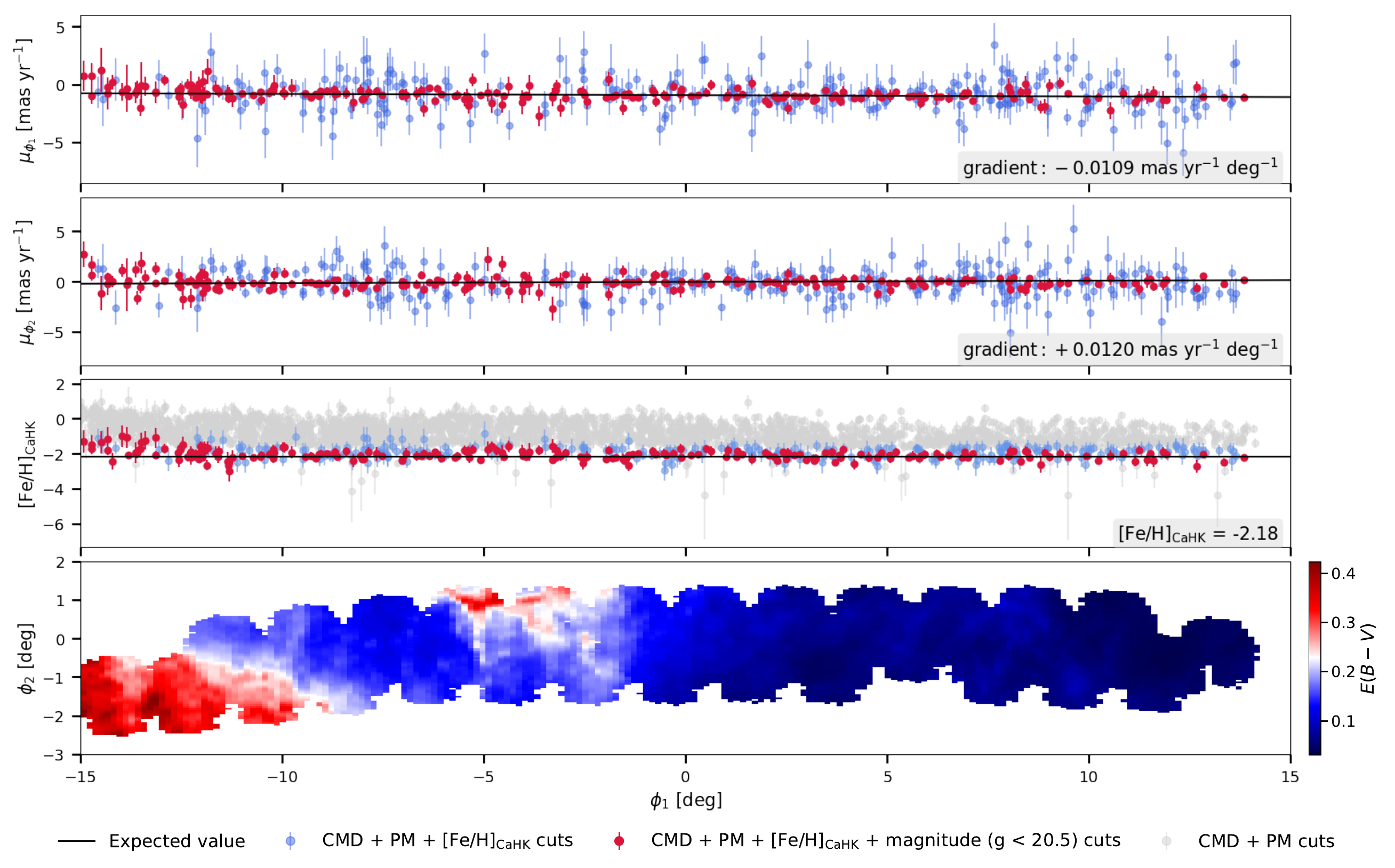}
\caption{Proper motion and metallicity measurements plotted against stream length. The first three panels in top-to-bottom order show the proper motion along $\phi_1$, proper motion along $\phi_2$, and metallicity ([Fe/H]$_{\text{CaHK}}$). The blue points represent all stars passing CMD, proper motion, and metallicity cuts, while red points represent stars that additionally have magnitudes above $g_0 = 20.5$. The black lines in these three plots represent expected values. The light gray points in the third panel are stars having passed only the CMD and proper motion cut, demonstrating the relatively higher metallicity population of background and foreground stars. The final plot shows the color excess across the extent of the stream, demonstrating how selection criteria may have been affected by dust reddening effects.}
\label{fig:phi1params}
\end{figure*}

The proper motion in both dimensions, plotted along $\phi_1$, are shown in the top and middle panels of Figure~\ref{fig:phi1params}. They are largely in agreement with the expected functions, though in regions where $\phi_1 < -11^\circ$ greater variations appear potentially caused by increased leakage due to greater foreground stellar density near the Milky Way disk. 

\subsection{Photometric Metallicity Selection}
\label{subsec:FeH}
Following \citet{taa+22}'s identification of the Jet stream’s metallicity, we initially perform a selection centered on [Fe/H]$_{\text{CaHK}} = -2.38$ using our derived photometric metallicities (the precisions of which are discussed in subsection~\ref{subsec:obs}). Rather than adopting a fixed-width window, we define a 2\,$\sigma$ selection for each star based on its individual metallicity uncertainty.
Specifically, we retain stars whose [Fe/H]$_{\text{CaHK}}$ lies within 2\,$\sigma$ from [Fe/H]$_{\text{CaHK}}=–2.38$, corresponding to a 95\% confidence interval. 
We exclude stars with photometric metallicity uncertainties exceeding 0.7\,dex to further ensure that stars with unreliable metallicities are not passing through our selection.
We note that this precision cut only affects stars at the faint-end of our sample ($g_0 > 20.5$).
We also exclude stars with obviously erroneous photometric metallicities, which include sources that appear beyond the metallicity grid in our catalog (i.e., [Fe/H]$_{\text{CaHK}} < -5.0$), which are likely spurious contaminants (e.g., galaxies,  variable sources), in addition to sources in the \textit{Gaia} DR3 variable source catalog \citep{gaiavar1, gaiavar2} and those that do not lie on the stellar locus of RGB stars in $r_0-i_0$ vs. $g_0-r_0$ space (e.g., Figure~3 in \citealt{apk+26}). 
After these steps, we find that the resulting metallicity distribution function peaks instead at [Fe/H]$_{\text{CaHK}} = -2.18$. We repeat the metallicity selection, revising the mean metallicity value to [Fe/H]$_{\text{CaHK}} = -2.18$, and adopt this as the metallicity of the Jet stream in our study. A 2\,$\sigma$ selection around [Fe/H]$_{\text{CaHK}} = -2.18$ yields 487 stars.
To explore the reasons for this difference, we compute the metallicity difference between our sample of candidate members and spectroscopic members from the $S^5$ survey. We find a mean zero-point offset in the metallicities between MAGIC and $S^5$ of $-0.11$ dex from our selection of Jet stream candidate stars that have $S^5$ measurements. This accounts for around half of the 0.2 dex discrepancy. The remaining discrepancy is likely due to minor foreground contamination increasing the mean metallicity of candidates. These metallicity considerations are further discussed in subsection~\ref{subsec:purity}.

The spatial distribution of stars passing this photometric metallicity cut, and the resulting distribution of metallicities, is shown in the bottom panels of Figure~\ref{fig:cuts}. We also show the variation of metallicity along $\phi_1$ in the bottom panel of Figure~\ref{fig:phi1params}. There appears an overall increase in the computed photometric metallicities in regions where $\phi_1 < -11^\circ$, potentially explained as an effect from the higher density of foreground stars closer to the Milky Way disk.

Further, we apply a magnitude cut selecting for stars $g_0 < 20.5$ in order to ensure small error values in \textit{Gaia} DR3 proper motions ($\lesssim0.5$ mas yr$^{-1}$). This narrows our selection of Jet stream candidates down to 213 stars.

\begin{deluxetable}{lccc}
\tablecaption{Adopted Jet stream Parameters \label{tab:params}}
\tablehead{
\colhead{Parameter} & \colhead{Value} & \colhead{Unit}  & \colhead{Source}
}
\startdata
Age\tablenotemark{$\dagger$} & 12.1 & Gyr & \citealt{jtn+18} \\
Photometric metallicity ([Fe/H]$_{\text{CaHK}}$)\tablenotemark{$\ddagger$} & $-2.18$ & \dex & See Section~\ref{subsec:FeH} \\
Spectroscopic metallicity ([Fe/H]) & $-2.38$ & \dex & \citealt{taa+22} \\
Distance modulus ($\mu$) & 17.55 & mag & See Section~\ref{subsec:cmd} \\
$\frac{\mathrm{d}\,d_\text{Jet}}{\mathrm{d}\,\phi_1}$ & $-0.2$ & kpc\,deg$^{-1}$ & \citealt{fsd+22} \\
$\mu_{\phi_1}$ & $-0.933 \pm 0.018$ & mas\,yr$^{-1}$ & \citealt{fsd+22} \\
$\mu_{\phi_2}$ & $-0.023 \pm 0.016$ & mas\,yr$^{-1}$ & \citealt{fsd+22} \\
$\mathrm{d}\mu_{\phi_1}/\mathrm{d}\phi_1$ & $-0.0109$ & mas\,yr$^{-1}$\,deg$^{-1}$ & \citealt{fsd+22} \\
$\mathrm{d}\mu_{\phi_2}/\mathrm{d}\phi_2$ & $+0.0120$ & mas\,yr$^{-1}$\,deg$^{-1}$ & \citealt{fsd+22} \\
\enddata
\tablenotetext{\dagger}{We note that this is slightly different from the age of the empirical isochrone used in this study (12.7\,Gyr; see Section~\ref{subsec:cmd}).}
\tablenotetext{\ddagger}{This is not a reported metallicity of the stream, but rather the metallicity selection centroid in this study. The selection criteria for candidate members is detailed in Section~\ref{sec:analysis}.}
\end{deluxetable}

\section{Results and Discussion}
\label{sec:discussion}

In the previous section, we described how we identified Jet stream member stars in the MAGIC catalog with the use of photometric metallicities. Below, we discuss the quality of our selection (Section~\ref{subsec:purity}), stream morphology as compared to previous investigations of the Jet stream (Section~\ref{subsec:morph}), as well as improvements for further analysis of the Jet stream (Section~\ref{subsec:future}).

\subsection{Purity and completeness of selection using photometric metallicities}
\label{subsec:purity}
An important test of the reliability of our analysis is to establish the purity of our sample of candidate Jet members. Purity is defined as the fraction of true stream member stars relative to all candidates in our final selection. 
We evaluate the purity of the Jet stream star sample by comparing it against observations of Jet stream stars made by $S^5$ (Ferguson et al., in prep).
$S^5$ uses the AAOmega spectrograph \citep{sss+06} to obtain medium-resolution ($R\sim$10000 over 8420\,{\AA} to 8820\,{\AA}) spectra of stars, from which metallicities and radial velocities can be derived.

$S^5$ derives membership probabilities for each star to be a member of the Jet stream using a two component Gaussian mixture model for the stream and foreground, similar to the methods described in Section 3.2 of \citet{slp+19} and Section 3.2 of \citet{ale+25}. Each group is described by Gaussian distributions in radial velocity, proper motion, and metallicity ([Fe/H]). The modeling allows the mean radial velocity and velocity dispersion to change along the stream as a function of $\phi_1$, and the mean proper motion to vary with $\phi_1$ with a fixed dispersion. The stream metallicity is modeled with a single Gaussian in [Fe/H] that has a fixed mean and dispersion for the whole stream. The same properties of foreground stars are described as Gaussians with values that do not vary with $\phi_1$ for all parameters. From this two–component Gaussian mixture model, the probability that each star is a member of the stream as opposed to the foreground is computed. 
To ensure a consistent comparison to $S^5$, we select from their catalog stars that are in spatial regions overlapping with our own, and eliminate BHB stars in their catalog as our selection of stars only includes those on the RGB and subgiant branch (see CMD in Figure~\ref{fig:footprint}).

We characterize the contamination in our sample by measuring, among all of our candidate members that are in the $S^5$ catalog, how many are clearly non-members (with $S^5$-provided membership probabilities $< 10\%$). We calculate this as the ratio of non-members to the total number of stars that pass a set of selection cuts in the cross-matched $S^5$ and MAGIC catalog. Using only a CMD-based selection, 957 of 1,034 stars present in both our and the $S^5$ catalog were identified as non-members, corresponding to a contamination of 92.6\%. Incorporating proper motion constraints reduced the fraction of non-members to 246 out of 318 stars (77.3\% contamination). Finally, adding a metallicity selection produced a further refined sample, where 13 of 71 stars were non-members, yielding a contamination of 18.3\% or a purity of 81.7\%. This purity remains virtually the same for looser constraints on membership probability, where a 50\% membership probability cut only changes the purity value by 2 stars.
Each additional cut markedly improves the proportion of stream members. This level of purity, while not yet sufficient to allow definitive conclusions regarding the presence or morphology of stream features from individual off-track stars, still results in a robust sample. The majority of retained stars will likely be members, providing a reliable basis for statistical analyses of the stream. Additionally, the sample will enable more efficient spectroscopic follow-up of candidate members.

In a similar manner, completeness is defined as the fraction of total stream stars that our selection is able to detect. We note that the completeness criteria relative to $S^5$ is likely imprecise as the $S^5$ catalog only has a subset of Jet members. 
In order to evaluate for completeness, we begin with 58 stars from the $S^5$ selection that have membership probabilities $>99\%$ and are consistent with the RGB of the Jet stream CMD.
A sequential proper motion cut reduced this number to 55, and a metallicity cut further reduced it to 51. 
A majority of stars on the red giant branch flagged as members of the Jet stream in the $S^5$ catalog are preserved through the selection process. 
We note that of the 213 candidate members that we identify, 71 have spectroscopy provided by $S^5$, with 142 requiring spectroscopic follow-up.

\subsection{Stream morphology}
\label{subsec:morph}
\begin{figure*}[!htbp]
\centering
\includegraphics[width =\textwidth]{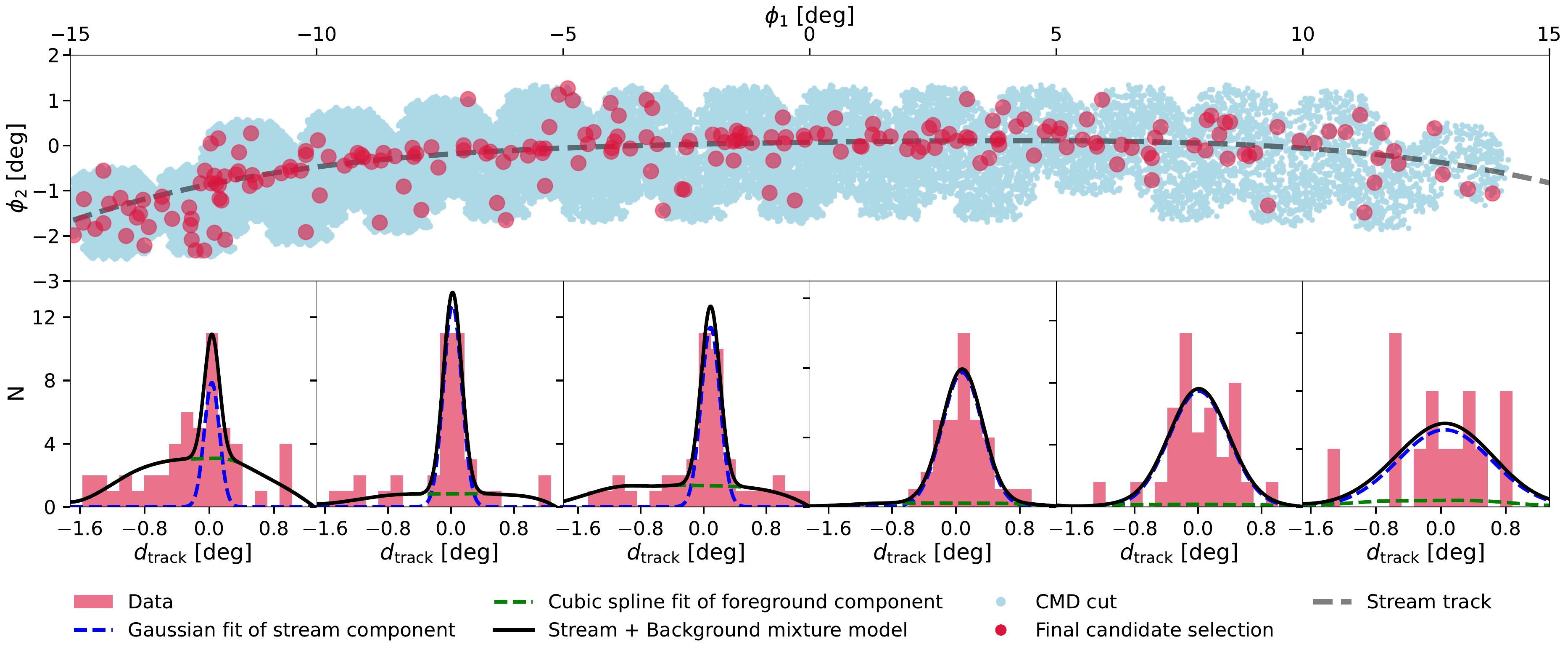}

\caption{Top: Spatial distribution of candidate stream stars in stream-aligned coordinates $\phi_1, \phi_2$. Blue points show stars selected by a CMD selection, and red points show the subset that additionally have proper motions and metallicities consistent with stream membership and magnitudes brighter than $g_0$ = 20.5. The dashed curve shows the stream track proposed in \citet{fsd+22}.
Bottom: Histograms of the perpendicular distances of final candidate stars to the stream track ($d_{\text{track}}$) in $5^\circ$ segments along $\phi_1$. The solid black curve is the best-fit model consisting of a Gaussian stream component plus a scaled spline model of the foreground, where the spline is derived from the distribution of all observed stars within a particular segment. The dashed blue line shows the Gaussian component alone and the dashed green line shows the spline foreground scaled by its fitted amplitude}
\label{fig:streamwhist}
\end{figure*}

\begin{figure}[!htbp]
\includegraphics[width =\columnwidth]{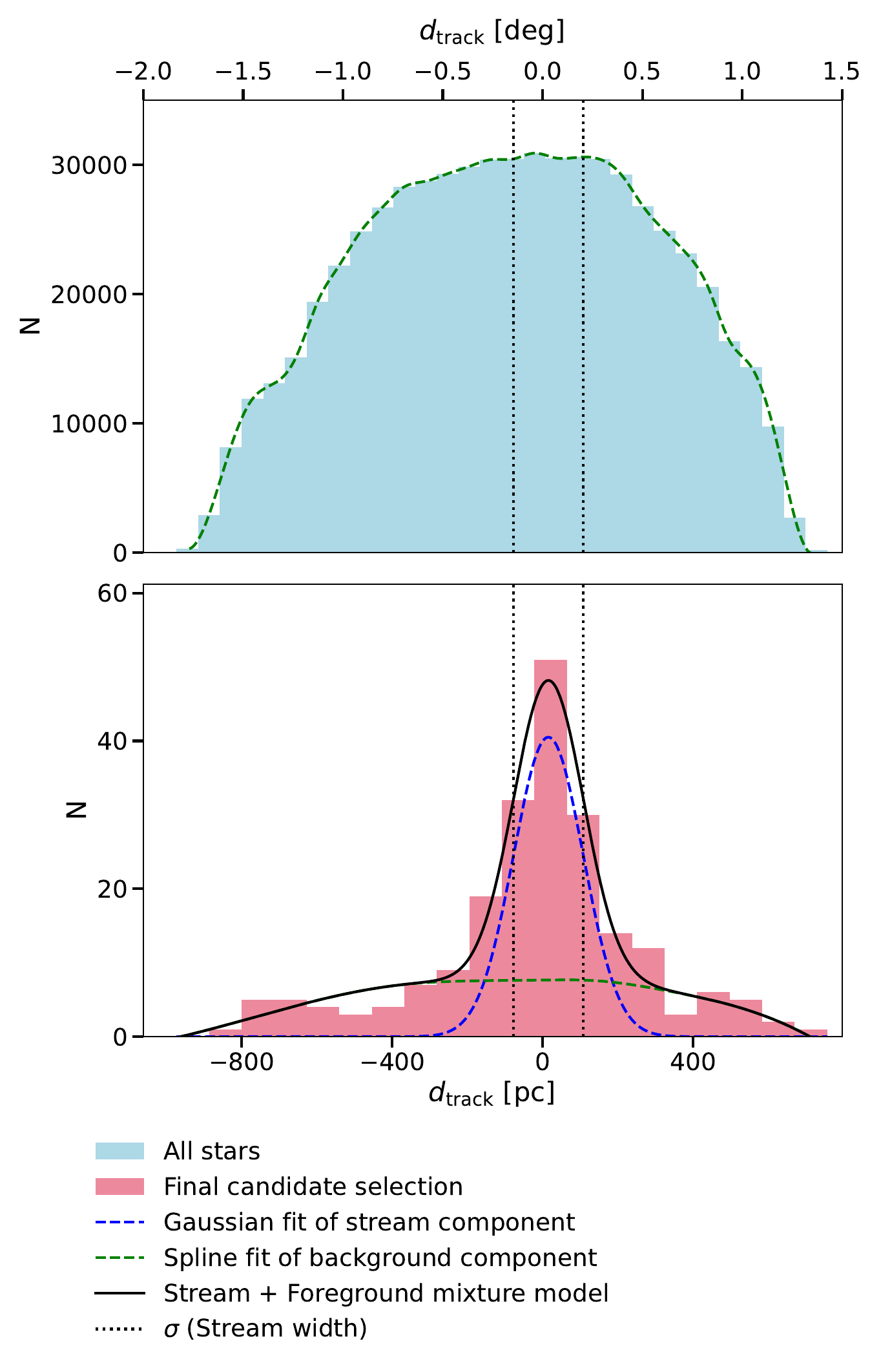}
\caption{Top: Histogram of the perpendicular distances of all observed stars from the stream track ($d_{\text{track}}$) and cubic spline constructed from this histogram, used to model the distribution of foreground Milky Way stars. Bottom: Histogram of the $d_{\text{track}}$ of Jet stream candidate stars, fitted with a two-component model consisting of a Gaussian representing the stream, and a scaled version of the previously-constructed spline to model foreground contamination. The overall model is shown by the black solid line, while the Gaussian and spline are blue and green dashed lines, respectively. Dashed vertical black lines indicate the $\sim 0.17^{\circ}$ width ($\sigma$) of the stream, translating to a physical width of $\sim 88$ pc.}
\label{fig:bkgspline}
\end{figure}

Visually in Figure~\ref{fig:streamwhist}, the stream density appears to vary along the stream. A narrow component of the stream is apparent between $\phi_1 \sim -10^\circ$ and $\phi_1 \sim 2.5^\circ$.
Toward larger $\phi_1$ ($> 7.5^\circ$), in the trailing region ($\phi_1 < -10^\circ$), and potentially around $\phi_1 \sim-5^\circ$, the stream appears to become more diffuse, with a broader spread in $\phi_2$. The broadening in regions of higher source density closer to the Milky Way disk ($\phi_1 < -10^\circ$, where $|b| \lesssim 9^\circ$) reflects greater reddening and/or leakage of foreground stars, though broadening outside this region may also be indicative of real perturbation features. 
The fanning of the stream at $\phi_1 > 0$ broadly corroborates the decrease in surface brightness and linear density that \citet{fsd+22} characterizes in the Jet stream around this region in Figure 6 of their paper.
Other features suggested by \citet{fsd+22}, such as the spur at $\phi_1 \sim 5^\circ$ and gap at $\phi_1 \sim - 6^\circ$ (though the latter could be the diffused region at $\phi_1\approx-5^\circ$ that we mention) are not recovered down to the magnitude depth of this study.
This is possibly due to an insufficient number of identified members and completeness/purity level in our study; alternatively, some of the suggested features in prior literature may not be associated with the stream.

We quantitatively evaluate the stream’s width using a two-component model (Gaussian and cubic spline) to represent the stream and foreground star components. We construct a model of the distribution of foreground stars that would arise from our pointings using the full sample of stars and computing their perpendicular distance to the stream track ($d_{\text{track}}$). We then form a histogram of these distances and fit a cubic spline to the binned counts as a smooth function of $d_{\text{track}}$, as shown in the top panel of Figure~\ref{fig:bkgspline}. Here, we use a cubic spline to account for the asymmetry in the shape of  each hexagonal pointing and in their distribution along the stream track, which does not allow them to be well described by a Gaussian or truncated Gaussian when binned along $\phi_2$. We then take the sample of candidate Jet stream members that pass our selections in Section~\ref{sec:analysis} and similarly compute $d_{\text{track}}$ for these stars. We fit a two-component model to the resulting histogram using a Gaussian to represent the detected stream and the previously-computed spline, scaled appropriately, as the foreground component:

\begin{equation}
N(d_{\text{track}}) = A \exp\!\left[-\frac{1}{2}\left(\frac{d_{\text{track}}}{\sigma}\right)^{2}\right] + B\,S{d_{\text{track}}}.
\end{equation}
Here, $N(d_{\text{track}})$ is the observed number density of stars as a function of $d_{\text{track}}$. $Sd_{\text{track}}$ is the spline-interpolated foreground component and $A$, $\mu$, $\sigma$, and $B$ are free parameters. The model is shown in the bottom panel of Figure~\ref{fig:bkgspline}.

In order to account for variation in the morphology of the stream along its length, we divide the stream into six equal segments along $\phi_1$ with limits: $(-15^\circ, -10^\circ)$, $(-10^\circ, -5^\circ)$, $(-5^\circ, 0^\circ)$, $(0^\circ, 5^\circ)$, $(5^\circ, 10^\circ)$, $(10^\circ, 15^\circ)$. For each segment, we repeat the spline construction process and two-component model fitting. We record the best-fitting amplitude, mean offset from the track, and standard deviation, along with the scaling factor of the foreground spline for each $5^\circ$ segment of the stream along $\phi_1$. The results of this modeling are shown in the bottom panel of Figure~\ref{fig:streamwhist} and Table~\ref{tab:segments}.

\begin{deluxetable}{lcccc}
\tablecaption{Gaussian-plus-spline fit results in bins of $\phi_1$
\label{tab:segments}}
\tablewidth{\columnwidth}
\tablehead{
\colhead{$\phi_1$ segment (deg)} &
\colhead{Amplitude} &
\colhead{Mean (deg)} &
\colhead{$\sigma$ (deg)} &
\colhead{Spline scale $B$}
}
\startdata
$[-15, -10]$ & $7.868$ & $0.0331$ & $0.0939$ & $0.708$ \\
$[-10, -5]$  & $12.760$ & $0.0204$ & $0.1164$ & $0.392$ \\
$[-5, 0]$    & $11.367$ & $0.0887$ & $0.1199$ & $1.429$ \\
$[0, 5]$     & $7.725$  & $0.0796$ & $0.2530$ & $0.338$ \\
$[5, 10]$    & $4.661$  & $0.0103$ & $0.3761$ & $0.252$ \\
$[10, 15]$   & $1.331$  & $0.0480$ & $0.5750$ & $0.546$ \\
\enddata
\end{deluxetable}

\begin{figure}[!htbp]
\includegraphics[width =\columnwidth]{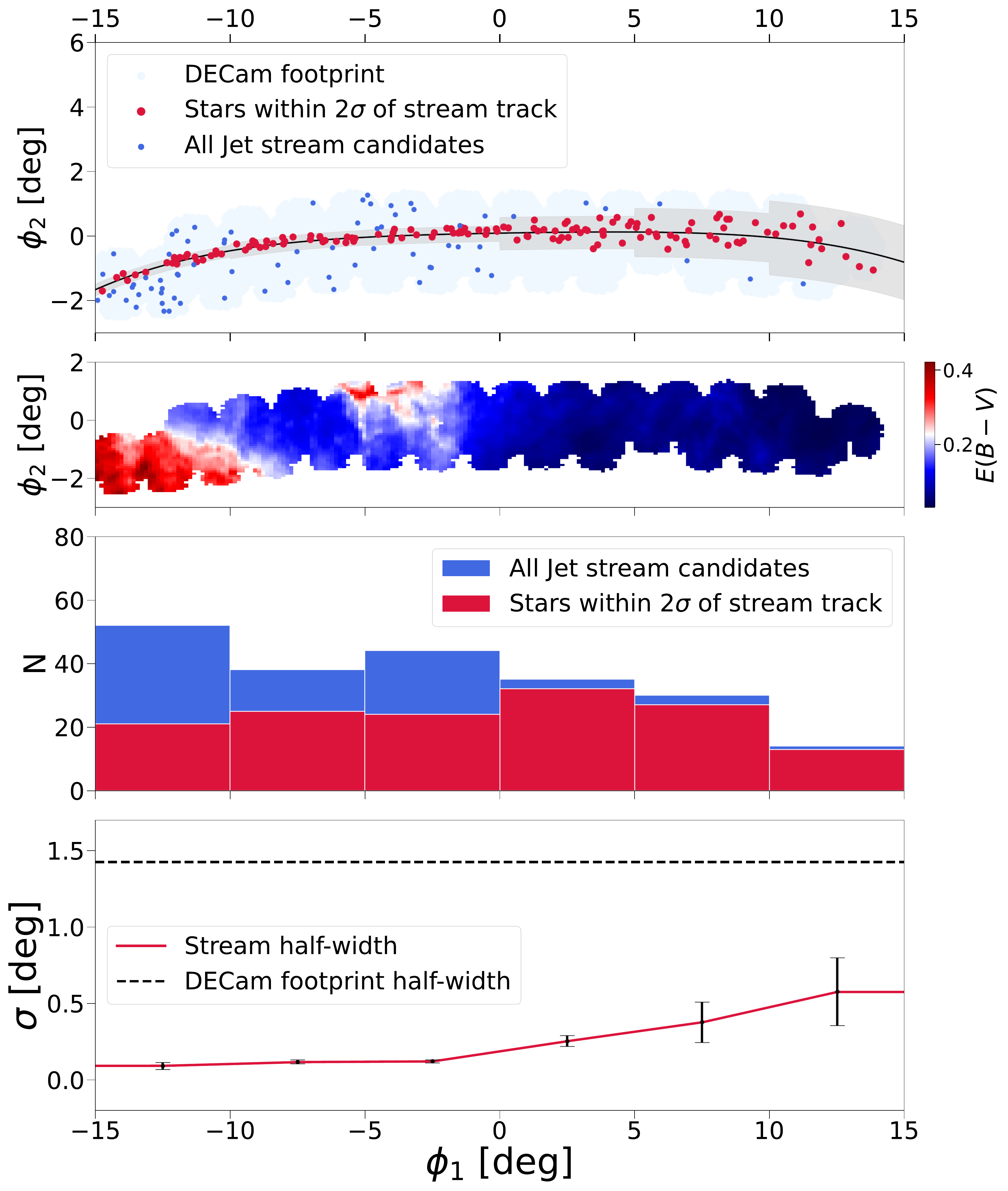}
\caption{First panel: stellar positions in the rotated coordinate system ($\phi_1$, $\phi_2$), compared to the best-fit stream track (black line). The shaded light–gray band shows the 2\,$\sigma$ envelope around the stream track, where $\sigma$ is the width of the $\phi_2$ residual distribution measured independently in the six $\phi_1$ bins in Figure~\ref{fig:streamwhist}. Stars lying within 2\,$\sigma$ of the track in their respective $\phi_1$ bin are plotted in red, while all other stars are shown in light blue. Second panel: $E(B-V)$ from \citet{sfd+98} across the extent of the stream, showing increased reddening in primarily the first and third 5$^\circ$ intervals of the stream along $\phi_1$. Third panel: Histogram of the $\phi_1$ distribution. The blue bars represent the entire sample of stars, and the red bars show only the subset of stars classified as being within 2\,$\sigma$ of the stream track. The bin boundaries match the $\phi_1$ intervals used to measure $\sigma$. Bottom: The red line shows stream width $\sigma$ as a function of $\phi_1$, demonstrating how width varies along the stream. The dashed black line shows the greatest distance between a single star and the stream track ($d_{\text{track}}$).}
\label{fig:dmorph}
\end{figure}

We find from this fitting that the overall width of the stream, computed as the standard deviation of the Gaussian in Figure~\ref{fig:bkgspline}, is $\sigma = (0.17 \pm 0.01)^{\circ}$. This is comparable to and thus corroborates \citet{fsd+22}'s estimate of $\sigma = 0.18^{\circ}$.
We further explore the stream width variation as a function of $\phi_1$ by investigating each $5^\circ$ segment. We plot the stream width in each segment along $\phi_1$, demonstrating the broadening of the stream width toward larger $\phi_1$ as is shown in the bottom panel of Figure~\ref{fig:dmorph}. 

This broadening of the Jet stream can potentially be interpreted in line with the mechanism suggested in \citet{wsj+16}. They provide mock models of the formation of the Ophiuchus stream, noted for its short visible portion ($\sim 1.5$ kpc) and high positional and kinematic dispersion of probable member stars. The authors suggest that the visible stream represents only the most recently disrupted material, while older material forms extended, diffuse “fans” with large dispersions in position and velocity, caused by chaotic evolution driven by the time-dependent, triaxial potential of the Galactic bar. In this regime, stars stripped at earlier close passages rapidly spread apart in phase space, forming faint, extended debris rather than remaining confined to a thin, dynamically cold filament. Further, \citet{yle+25} suggest that interactions with Galactic bar  leads to a ``flipping" of the tidal tails in Ophiuchus, where the fanning is from either leading or trailing tails released at different time intervals. We note that the galactocentric radius of the Jet stream is over an order of magnitude greater than that of Ophiuchus (30.4 and 1.5 kpc respectively); more detailed modeling would be needed to conclusively explain the fanning seen in the Jet stream.

As a separate assessment of purity, we select stars whose $d_{\text{track}}$ lies within 2\,$\sigma$ of the expected stream track (red points in the top panel of Figure~\ref{fig:dmorph}) from candidate stars using the individually fitted widths of each segment. These stars make up $\sim67\,\%$ of the 213 candidate stars, empirically indicating a purity of this value if the stream's morphology were well-described by these segmented Gaussian fits.
If we were to limit the study to $\phi_1 > -10^{\circ}$, to exclude regions with high reddening approaching the Milky Way disk, this percentage increases to $\sim76\,\%$, similar to our purity estimate based on comparison to the $S^5$ spectroscopic sample.

\subsection{Future investigations}
\label{subsec:future}
In the near future, the Vera C. Rubin Observatory’s Legacy Survey of Space and Time (LSST; \citealt{ikt+19}) will provide significantly deeper photometric observations by several magnitudes, allowing the potential to resolve finer substructure and density variations within the Jet stream. The LSST $u-$band filter is metallicity-sensitive (e.g., \citealt{izk+19}, \citealt{pdi+25}), permitting the application of a photometric metallicity selection analogous to that employed in this work. The expected $5\sigma$ point-source depths of LSST are $u \simeq 26.1$ for coadded images idealized for stationary sources after 10 years \citep{biz+22}. LSST also potentially provides proper motion values with accuracies expected to be $0.2$\,mas/yr for objects as bright as $r = 21$ and $1\,$mas/yr for objects down to $r = 24$ \citep{ibj+12}, whereas the current proper motion values from \textit{Gaia} DR3 retain an uncertainty of $0.5$\,mas/yr for objects at $G = 20$ \citep{lkh+21}. 
As an illustrative example, we compute the improvement in number of stars that a depth gain of two magnitudes will achieve. 
We integrate the luminosity function over the increased magnitude range of $g_0 = 20.5$ to $g_0 = 22.5$ using an isochrone with an age of 12.1 Gyr and [Fe/H] = $-2.18$ interpolated from the Dartmouth Stellar Evolution Database \citep{dcj+08}, which approximately corresponds to the parameters of the Jet stream. 
This results in a $\sim12\times$ increase in the number of potential stream members that can be probed. Based on our number of 213 candidate members of the Jet stream down to $g_0 \sim20.5$, we forecast that the improved photometric depth of Rubin/LSST will yield $\sim$2,700 members along the Jet stream. This will enable deeper investigation of the stream’s faint structures, expanding on our analysis in this work.
However, this should be viewed as an optimistic scenario as the metallicity performance for the LSST $u$ filter for warmer stars at the main-sequence turn-off and along the main-sequence will likely degrade relative to cooler RGB stars. 
We note that more immediate improvements are likely possible with \textit{Gaia} DR4 and DR5, which will provide improved proper motion precisions by $\sim3\times$ and $\sim7\times$ relative to \textit{Gaia} DR3 \citep{brown+25}. 

Conclusive confirmation of stream membership will also require spectroscopic follow-up. We therefore provide a table of candidate members in Appendix \ref{app:cat} to enable both membership confirmation and further kinematic analysis of stars that deviate from the stream track.
In particular, radial velocities of off-track stars can be used to confirm their association with the stream, and the variations in the velocity dispersion along the stream track can be used to assess the dynamical state of the stream.

\section{Conclusion}
\label{sec:conclusion}

We identified stars that are likely members of the Jet stellar stream using a combination of photometric metallicities from DECam narrow-band CaHK imaging, combined with \textit{Gaia} DR3 proper motions and broadband imaging data. We validate our selection against $S^5$ spectroscopy, showing that sequential CMD, proper-motion, and metallicity cuts allow for the effective identification of Jet stream candidate members, yielding a sample suitable for morphological analysis and future spectroscopic follow-up. Our results confirm that wide-field photometric metallicities and astrometric data can be used for isolating members of faint stellar structures, such as the Jet stream. We summarize our takeaways:
\begin{itemize}
    \item We clearly recover the Jet stream (Figure~\ref{fig:cuts}), and estimate purity of $\sim 82\%$, based on comparison to spectroscopically confirmed Jet stream members from the $S^5$ survey (Ferguson et al., in prep).
    \item We demonstrate that our photometric metallicities, when combined with astrometry, can be used to reliably isolate a clean sample of Jet stream member stars. We find that photometric metallicities can reduce contamination from non-stream stars from $\sim$76\% to $\sim$18\%, yielding a significantly cleaner Jet stream member sample than CMD and proper-motion cuts alone.
    \item We corroborate previous measurements of the stream width ($\sigma\sim0.18^{\circ}$; \citealt{fsd+22}), and find evidence of significant stream broadening where $\phi_1 > 5^\circ$.
    \item Some previously suggested morphological features such as the spur at $\phi_1 \sim 5^\circ$, potentially artefacts of interaction with dark matter subhalos, are not clearly recovered in this analysis, motivating further investigation with deeper imaging and kinematic data.
    \item Visually, we identify a region of slightly diffused density around $\phi_1 \sim -5^\circ$ and a broadening of the stream when $\phi_1\gtrsim 2.5^\circ$, corroborating aspects of the morphology suggested in \citet{fsd+22}, suggesting the possibility of real stream features.
    \item We provide a catalog of potential Jet stream members in Appendix \ref{app:cat}, including the data used in their selection, in order to enable future follow-up. This table contains our 213 candidate members, 71 of which have good spectroscopic data from $S^5$.
    
\end{itemize}

This study, together with other recent analyses of nearby dwarf galaxies, stellar clusters, and substructures \citep[e.g.,][]{longeard+22, martin+22, ash+22, fu+23, kuzma+25, bcl+25, apk+26}, has demonstrated the utility of using narrow-band photometry survey data in stream and satellite characterization. In particular, these previous studies show that the use of the narrow-band CaHK filter and its derived metallicities has proven effective both for identifying likely members and for characterizing the chemical compositions of stellar systems. With the planned coverage of $\sim 5000~\mathrm{deg}^2$, the MAGIC survey allows for these methods of analysis to be scaled for the study of other stellar streams across the local Milky Way halo. Forthcoming LSST data will add several magnitudes of depth and deliver proper motions for fainter stars, enabling the characterization of a larger volume of stars and low-surface-brightness structure.

\begin{acknowledgements}

We acknowledge The College at the University of Chicago for their support of undergraduate research through the Quad research fellowship, which allowed for this analysis of the Jet stream. A.C. is supported by a Brinson Prize Fellowship grant through the Brinson Foundation. W.C. gratefully acknowledges support from a Gruber Science Fellowship at Yale University. This material is based upon work supported by the National Science Foundation Graduate Research Fellowship Program under Grant No. DGE2139841. Any opinions, findings, and conclusions or recommendations expressed in this material are those of the author(s) and do not necessarily reflect the views of the National Science Foundation. S.K. acknowledges support from Science \& Technology Facilities Council (STFC) (grant ST/Y001001/1). G.F.L. acknowledges support from the Australian Research Council through the Discovery Program grant DP220102254. DJS acknowledges support from NSF grants AST-2205863 and AST-2508746. D. B. Z. acknowledges support from the Australian Research Council through the Discovery Program grant DP220102254.

The DELVE project is partially supported by the NASA Fermi Guest Investigator Program Cycle 9 No. 91201.
This work is partially supported by Fermilab LDRD project L2019-011. 
This material is based upon work supported by the National Science Foundation under Grant No. AST-2108168, AST-2108169, AST-2307126, and AST-2407526.
This project used data obtained with the Dark Energy Camera (DECam), which was constructed by the Dark Energy Survey (DES) collaboration. 
Funding for the DES Projects has been provided by the US Department of Energy, the U.S. National Science Foundation, the Ministry of Science and Education of Spain, the Science and Technology Facilities Council of the United Kingdom, the Higher Education Funding Council for England, the National Center for Supercomputing Applications at the University of Illinois at Urbana–Champaign, the Kavli Institute for Cosmological Physics at the University of Chicago, the Center for Cosmology and Astro-Particle Physics at the Ohio State University, the Mitchell Institute for Fundamental Physics and Astronomy at Texas A\&M University, Financiadora de Estudos e Projetos, Fundação Carlos Chagas Filho de Amparo à Pesquisa do Estado do Rio de Janeiro, Conselho 12 Nacional de Desenvolvimento Científico e Tecnológico and the Ministério da Ciência, Tecnologia e Inovação, the Deutsche Forschungsgemeinschaft and the Collaborating Institutions in the Dark Energy Survey.

The Collaborating Institutions are Argonne National Laboratory, the University of California at Santa Cruz, the University of Cambridge, Centro de Investigaciones Enérgeticas, Medioambientales y Tecnológicas–Madrid, the University of Chicago, University College London, the DES-Brazil Consortium, the University of Edinburgh, the Eidgenössische Technische Hochschule (ETH) Zürich, Fermi National Accelerator Laboratory, the University of Illinois at Urbana-Champaign, the Institut de Ciències de l’Espai (IEEC/CSIC), the Institut de Física d’Altes Energies, Lawrence Berkeley National Laboratory, the Ludwig-Maximilians Universität München and the associated Excellence Cluster Universe, the University of Michigan, NSF NOIRLab, the University of Nottingham, the Ohio State University, the OzDES Membership Consortium, the University of Pennsylvania, the University of Portsmouth, SLAC National Accelerator Laboratory, Stanford University, the University of Sussex, and Texas A\&M University.

Based on observations at NSF Cerro Tololo Inter-American Observatory, NSF NOIRLab (NOIRLab Prop. ID 2019A-0305; PI: Alex Drlica-Wagner, NOIRLab Prop. ID 2023B-646244; PI: Anirudh Chiti, NOIRLab Prop. ID 2025A-568024; co-PI: Anirudh Chiti, Ha Do), which is managed by the Association of Universities for Research in Astronomy (AURA) under a cooperative agreement with the U.S. National Science Foundation.

This manuscript has been authored by Fermi Research Alliance, LLC under Contract No. DE-AC02-07CH11359 with the U.S. Department of Energy, Office of Science, Office of High Energy Physics. The United States Government retains and the publisher, by accepting the article for publication, acknowledges that the United States Government retains a non-exclusive, paid-up, irrevocable, world-wide license to publish or reproduce the published form of this manuscript, or allow others to do so, for United States Government purposes.

This work has made use of data from the European Space Agency (ESA) mission {\it Gaia} (\url{https://www.cosmos.esa.int/gaia}), processed by the {\it Gaia} Data Processing and Analysis Consortium (DPAC, \url{https://www.cosmos.esa.int/web/gaia/dpac/consortium}). Funding for the DPAC has been provided by national institutions, in particular the institutions participating in the {\it Gaia} Multilateral Agreement.

\end{acknowledgements}

\software{Jupyter \citep{jupyter}, TOPCAT (\url{http://www.starlink.ac.uk/topcat/}, \citealp{topcat}), Astropy \citep{astropy:2013, astropy:2018, astropy:2022}, NumPy \citep{numpy:2020}, SciPy \citep{scipy:2020}, matplotlib \citep{matplotlib:2007}}

\facilities{Blanco (DECam), \textit{Gaia}}

\bibliography{references}{}
\bibliographystyle{aasjournal}

\appendix
\section{Catalog}
\label{app:cat}
\begin{table*}[ht]
\centering
\caption{Sample of Jet stream candidate members identified in this study. \textit{Gaia} Source ID, $\alpha$, $\delta$, proper motions, dereddened CaHK magnitudes from DECam observations, dereddened DELVE DR2 $g$ and $i$ band magnitudes, as well as photometric metallicities and corresponding errors are provided. The final column indicates whether spectroscopy from $S^5$ is available. The full table of candidate members, as well as full table of observed stars from the aggregate survey, are made available online \citep{zenodo}.}
\begin{tabular}{ccccccccccc}
\hline
  \texttt{Gaia DR3 Source ID} & \texttt{$\alpha$} & \texttt{$\delta$} & \texttt{$\mu_\alpha$} & \texttt{$\mu_\delta$} & \texttt{CaHK\_0} & \texttt{g\_0} & \texttt{i\_0} & \texttt{[Fe/H]} & \texttt{e\_[Fe/H]} & \texttt{spec.avail.$^\dagger$} \\
  \texttt{} & \texttt{(deg.)} & \texttt{(deg.)} & \texttt{(mas/yr)} & \texttt{(mas/yr)} & & & & & & \\
\hline \hline
  5682064161481234688 & 143.020166 & -16.740816 & -0.92 & -1.77 & 20.76 & 20.15 & 19.58 & -2.07 & 0.21 & No\\
  5682102408165005696 & 142.828520 & -16.863205 & -0.89 & -1.60 & 18.91 & 18.09 & 17.32 & -2.20 & 0.16 & Yes\\
  5681991941606439680 & 141.925121 & -17.091920 & -0.49 & -1.81 & 20.44 & 19.80 & 19.21 & -2.11 & 0.27 & No\\
  5682017878913194752 & 142.461754 & -17.265087 & -0.91 & -1.46 & 20.30 & 19.67 & 19.04 & -2.28 & 0.20 & No\\
  5682014344155793920 & 142.547795 & -17.282824 & -1.03 & -2.14 & 20.67 & 20.03 & 19.42 & -2.12 & 0.19 & No\\
  5682030557656916736 & 143.132500 & -17.190452 & -0.86 & -1.66 & 21.02 & 20.45 & 19.93 & -2.00 & 0.22 & No\\
  5681921224969378816 & 142.237013 & -17.435905 & -0.84 & -1.61 & 18.46 & 17.60 & 16.78 & -2.23 & 0.17 & No\\
  5681928925846184448 & 141.566300 & -17.791305 & -0.90 & -1.62 & 18.64 & 17.77 & 17.00 & -1.96 & 0.16 & Yes\\
  5681927237923304832 & 141.833958 & -17.735671 & -0.61 & -1.68 & 19.20 & 18.47 & 17.79 & -2.07 & 0.17 & No\\
  5681548421808268928 & 141.608429 & -17.938316 & 0.33 & -2.15 & 20.79 & 20.26 & 19.74 & -2.22 & 0.24 & No\\
  5681871201485633024 & 142.186078 & -17.789197 & -0.46 & -1.37 & 20.61 & 19.98 & 19.41 & -1.90 & 0.21 & No\\
  5681937717643590656 & 141.743298 & -17.638384 & -0.05 & -2.09 & 20.90 & 20.34 & 19.78 & -2.28 & 0.26 & No\\
  5679604176013055232 & 139.223069 & -20.601237 & -0.20 & -1.03 & 20.27 & 19.60 & 19.01 & -1.88 & 0.21 & No\\
  5676595706040611200 & 139.420597 & -20.767536 & -0.90 & -1.42 & 19.48 & 18.77 & 18.08 & -2.15 & 0.18 & Yes\\
  5679190106804793984 & 138.616059 & -21.134867 & -0.98 & -0.86 & 20.69 & 20.04 & 19.49 & -1.73 & 0.23 & No\\
  5676200465970655744 & 139.465789 & -21.116609 & -0.79 & -1.59 & 19.11 & 18.36 & 17.63 & -2.17 & 0.17 & No\\
  5676200397251198080 & 139.493363 & -21.095619 & -0.86 & -1.27 & 21.02 & 20.44 & 19.94 & -1.79 & 0.26 & No\\
  5676160196356213888 & 138.793589 & -21.514765 & -0.56 & -1.49 & 19.73 & 19.02 & 18.37 & -2.02 & 0.18 & Yes\\
  5676114326106157952 & 139.317600 & -21.490908 & 0.03 & -0.83 & 21.02 & 20.39 & 19.84 & -1.86 & 0.27 & No\\
  5676065054240878464 & 138.654580 & -21.635170 & -0.68 & -1.49 & 18.63 & 17.77 & 16.98 & -2.10 & 0.16 & Yes\\
  5676050623150445312 & 138.522009 & -21.876372 & -0.00 & -1.86 & 21.10 & 20.46 & 19.92 & -1.70 & 0.27 & No\\
  5676050863668553728 & 138.610069 & -21.908172 & -0.37 & -0.99 & 20.22 & 19.59 & 18.97 & -2.20 & 0.21 & Yes\\
  5690829399457859072 & 146.304668 & -12.478817 & -0.86 & -2.15 & 21.00 & 20.47 & 19.99 & -1.90 & 0.28 & No\\
  5690782395335674624 & 146.594266 & -12.577417 & -1.16 & -1.92 & 20.14 & 19.48 & 18.86 & -2.03 & 0.19 & Yes\\
  5690748482273641728 & 146.249069 & -12.831259 & -1.17 & -1.72 & 19.91 & 19.24 & 18.64 & -2.00 & 0.18 & Yes\\
\hline \hline
\end{tabular}
\tablenotetext{\dagger}{Only stars with good spectroscopically-derived quantities in the S$^5$ catalog (i.e., with \texttt{good\_star == True}) are indicated.}
\label{tab:cat}
\end{table*}

\end{document}